# Generating Rhythm Game Music with Jukebox

Nicholas Yan[1*]

[1]Farragut High School, Knoxville, Tennessee, USA

**\* Correspondence:**
Nicholas Yan
cubeymath@gmail.com

**Abstract**

Music has always been thought of as a "human" endeavor- when praising a piece of music, we emphasize the composer's creativity and the emotions the music invokes. Because music also heavily relies on patterns and repetition in the form of recurring melodic themes and chord progressions, artificial intelligence has increasingly been able to replicate music in a human-like fashion. This research investigated the capabilities of Jukebox, an open-source commercially available neural network, to accurately replicate two genres of music often found in rhythm games, artcore and orchestral. A Google Colab notebook provided the computational resources necessary to sample and extend a total of sixteen piano arrangements of both genres. A survey containing selected samples was distributed to a local youth orchestra to gauge people's perceptions of the musicality of AI and human-generated music. Even though humans preferred human-generated music, Jukebox's slightly high rating showed that it was somewhat capable at mimicking the styles of both genres. Despite limitations of Jukebox only using raw audio and a relatively small sample size, it shows promise for the future of AI as a collaborative tool in music production.

## 1 Introduction

### 1.1 Musical Genres Studied

The music of interest in this study belongs to two genres: orchestral and artcore. As the name suggests, orchestral music is traditionally performed by an orchestra. It includes a wide variety of instruments, including the piano, strings, woodwinds, brass, and percussion. Furthermore, the music has a more melodic quality and a tempo that varies significantly throughout different works, though selected works of this genre are slower. Songs may also feature electronic drums as accompaniment while keeping the orchestral arrangement as the focal point.

According to Reichert (2022), artcore is a subgenre of electronic music which incorporates acoustic instruments, primarily piano and strings, with electronic elements backed by a drum 'n' bass or breakbeat drumline. The music is typically aggressive in tone and has a tempo of at least 170 beats per minute (BPM). Additionally, certain electronic elements (atonal synths and white noise) cannot be transcribed into MIDI. People may differ in how they represent these sounds (either by using glissandi- sliding between the desired pitches- or omitting the section entirely), which hinders accurate recreation.

Both genres of music can be readily found in Deemo, a rhythm game released by Rayark Inc. in 2013. Each level is keysounded; in other words, every note in a level has a pre-programmed sound, such as a single piano note. This leads to greater player immersion and a significantly higher proportion of melodic songs compared to other rhythm games, which makes Deemo a good resource to gather songs for training.

**1.2     Overview of Particular Neural Network Models**

A neural network is a computing system modeled after a natural brain. Neural networks are composed of layers, with each layer having many nodes. The node is the smallest unit of a neural network and usually performs a simple operation. Usually, this means taking a weighted sum of the inputs, adding a constant (bias), and normalizing the result between a certain interval. This is shown in Figure 1 with an input layer drawn in yellow and nodes drawn in purple.

Values of nodes in the "hidden" middle layers are unknown to the user, establishing neural networks as "black boxes"- they produce useful outputs, though their internal workings are unknown. Lastly, neural networks become more efficient and accurate when continuously trained on a certain set of data. This impressive learning speed enables them to create unique data that closely resembles the inputted training data. Researchers have taken advantage of this by training neural networks on MIDI (musical instrument digital interface) files, which concisely convey information such as the notes' timing, loudness, and pitch.

**Figure 1**
*Basic representation of a neural network*

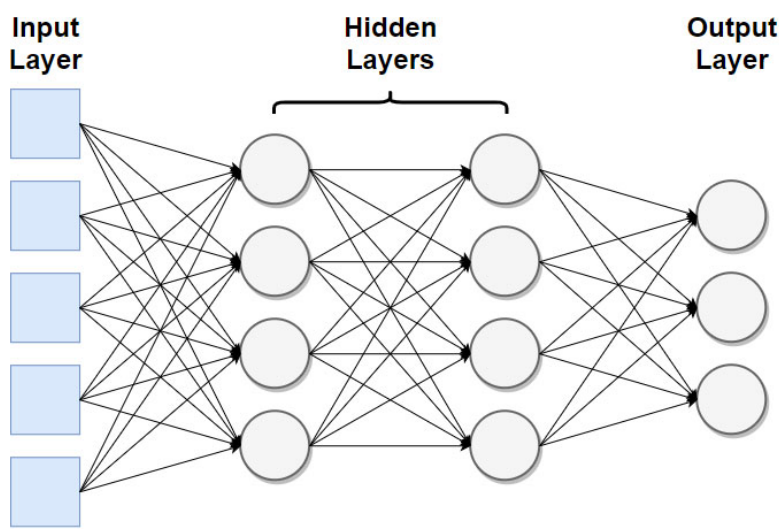

There are a few major neural network models which have been used for a wide range of applications, starting with the RNN (recurrent neural network). Johnson (2017) states that in a recurrent neural network, nodes can connect to other nodes in the next layer but also to themselves. This gives the neural network a rudimentary form of memory; with minimal changes, it can now remember its outputs one time step in the past. RNNs are often used with long short-term memory (LSTM). "Memory nodes," specifically designed to hold information from past time steps, further extend the effective memory of an RNN. A basic representation of an RNN is shown in Figure 2. The vertical axis follows the RNN through time while the horizontal axis follows the RNN through space. The memory node is shown in orange.

**Figure 2**

*Basic representation of an RNN throughout time using LSTM*

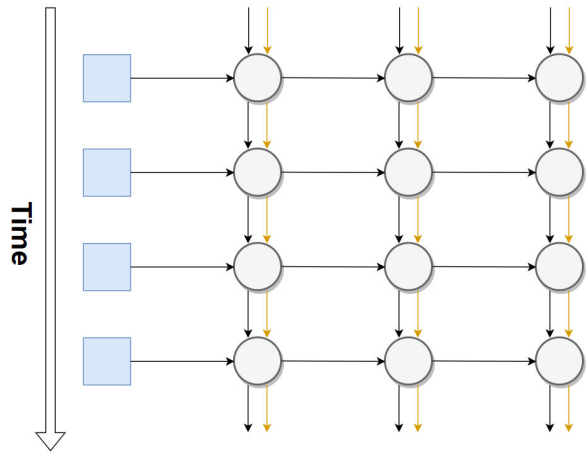

     Other neural network models used in music generation but are less relevant to the upcoming research are as follows: convolutional neural network (CNN) and generative adversarial network (GAN). According to O'Shea and Nash (2015), a CNN is a neural network that specializes in extrapolating patterns from *images*; researchers have repurposed CNNs by converting MIDI data to images (e.g., showing notes pressed as white against a black background). A GAN is a neural network composed of two parts: a generator and a discriminator. The generator produces data and the discriminator gauges its accuracy before feeding the data back into the generator, creating a positive feedback loop. Cao et al. (2020) have taken advantage of this model by using separate generators and discriminators for melodies and chords, expediting the neural network's training time.

**1.3    Examples of Established Neural Network Models**

     Overall, researchers have focused on using one of the neural network models described in the previous section and adapted it to their needs. This can be done by changing the architecture. Researchers may add or remove layers depending on the number of instruments to analyze and recreate, include or exclude discriminators, or create a new neural network model entirely. Several examples are given below with their corresponding model(s) and the type of music emulated.

     Chen et al. (2020) propose a GAN model to overcome the challenges of composing pop music, which include diversity in style and multi-track arrangement (usage of many instruments including the piano, drum, bass, violin, etc.) The proposed architecture has two discriminators: one to guide the style and another to confirm the harmony of the results. This combined model, the Multi-Style Multi-Instrument Co-Arrangement Model (MSMICA), is able to account for factors such as singability and arrangement, which other models cannot. Overall, the model has three parts: first, the data is processed and separated into its components (melody, chord progression, rhythm, instruments); second, the chord progression and rhythm of each instrument are analyzed; third, the instruments are put together. The results are fed through the style and harmony discriminators to produce new outputs before all steps are repeated as necessary. To ensure consistency in training, the parameters of the model were the same throughout all trials and the outputs were normalized to have the same tempo and key, reducing listener bias. Compared to a default GAN structure, MSMICA qualitatively scored 0.5 points higher from a scale from 1 to 5 in terms of harmony and style.

     In another approach, Li, Jang, and Sung (2019) introduce a GAN that is enhanced with two discriminators: LSTM (ensuring consistency between bars) and CNN (validating the overall musical

structure.) First, the GAN pre-processes the data by extracting the melody and encoding it into a matrix with dimensions based on pitch and time. To generate new MIDI, random noise was fed through the GAN and the discriminators rather than music of a certain genre. The efficacy of the GAN was measured using TFIDF (Term Frequency-Inverse Document Frequency). While is mainly used to measure the relevance of a keyword in a document, it could also be used to see the structural differences between the generated and real melodies by representing MIDI as alphanumeric characters. The TFIDF of the proposed model was 0.049 while that of real music was 0.053, meaning that the generated music showed diversity and harmony similar to that of real music.

Kaliakatsos-Papakostas et al. (2018) integrate LSTM in various neural network structures, including RNNs. The researchers avoided pre-processing by selecting educational piano pieces, which all had similar metadata (C major, 4/4, 120 bpm), musical diversity, and a limited range of 2 octaves. To transform the MIDI, two aspects were considered: sparsity (how often notes are played) and register (the pitches). Randomizing these values through user input and a random seed provided enough variation to prevent overfitting. Once this was done, the researchers tested the efficacy of the model using three different learning curves; slow- easing from training data into generating original music; fast- generating original music right away; and training- reusing training data in a different arrangement.

The three examples shown above are a small representation of the current literature within the field of music generation. There are still difficulties; neural networks cannot extrapolate musical patterns between long distances of time, and their results are not as coherent as their human counterparts. Nevertheless, these examples show promise, as neural networks are increasingly able to generalize certain genres of music.

**1.4    Jukebox**

Jukebox is a generative model that creates music with or without lyrics as raw audio. In other words, it outputs data in .wav format instead of MIDI. According to Dhariwal et al. (2020), Jukebox uses an auto-encoder based on a CNN to compress, encode, decode, and reconstruct audio on three different temporal resolutions, ensuring that the neural network is able to recognize patterns across multiple reference frames. It can either generate completely new music or extend a given sample; the following research will focus on the second feature. To do this, Jukebox samples from a certain time window at each time step, ensuring a smooth transition between the input and generated audio. This flexibility is further augmented by a training database of over 1.2 million songs of various genres, over 600 thousand of which have lyrics.

**1.5    Gap in Research**

Due to Jukebox's reputation as a reliable and powerful tool for generative music and its ease of use, it will be used for conducting this study. As seen from the examples above, the main musical genres that other research has focused on have been classical (Romantic, Baroque, etc.), pop, or folk. There are reliable databases with thousands of accurate MIDI files available in these genres. On the other hand, there is no specific database for the rhythm game-style music described earlier. Artcore is relatively recent, only achieving status as a niche subgenre in the 2000s, and as stated previously, atonal electronic elements prevent accurate transcription. However, through careful song selection, this issue can be mitigated and allow Jukebox to produce purely melodic music in a similar style. As previous research has not focused on these niche subgenres of music, this raises the question: According to local youth orchestra members, how accurately can Jukebox, a convolutional neural network, create harmonious music by extending melodies of artcore and orchestral songs?

## 2    Methods

A neural network was needed to generate songs of both genres- artcore and orchestral. A Google Colab notebook running Jukebox was the ideal choice, as it gave the user temporary access to a Google-owned GPU and its fill-in-the-box interface allowed for easy use as compared to coding a neural network from scratch. Google Colab is also free, albeit with restrictions on usage hours. Figure 3 shows the Google Colab notebook parameters with sample inputs. As seen below, Jukebox is instructed to extend the audio file "Oceanus.mp3" using the first 30 seconds as a prompt for an additional 60 seconds in a classical style.

**Figure 3**
*Screenshot of Jukebox parameters*

```
model: 5b
hps.n_samples: 2
hps.name: "/content/gdrive/MyDrive/Project_4"
speed_upsampling: ✓
mode: primed
audio_file: "/content/gdrive/MyDrive/Oceanus.wav"
prompt_length_in_seconds: 30
sample_length_in_seconds: 90
select_artist:
select_genre: classical
sampling_temperature: .98
disconnect_runtime_after_finish: ✓
```

The parameters inputted into the Jukebox Colab notebook were kept consistent when possible. I used the most recent model at the time of writing (5b) without support for lyrics, as they were unnecessary, and the samples were generated one at a time (hps.n_samples) due to my laptop's computational constraints. For each excerpt generated, the first 30 seconds of each song were used as a prompt. Jukebox then used the prompt as a reference to extend the song for 60 seconds, for 90 seconds of audio in total. This let the original song sufficiently develop while ensuring enough time for the AI to create new material. To avoid the inclusion of vocals and other non-piano instruments, the genre "classical" was used. Lastly, the sampling temperature was set at 0.98 to avoid overfitting, or the regurgitation of training data. This ensured sufficient originality of the neural network.

A total of 16 songs were used as input. As I was unable to obtain the original MIDI data due to copyright, piano arrangements of these songs were chosen instead. All arrangements had to be physically playable, which required the usage of a left and right hand. The time signature was kept constant, and the tempo was kept constant within ten seconds of the 30-second mark, allowing Jukebox sufficient time to adjust to the new tempo. Arrangements were also chosen based on the arranger's credibility; the most prominent arrangers below (phyxinon, Ayato Fujiwara, Salamanz) had received positive feedback from the Deemo community over years in the form of views, likes, and comments. If none of the arrangers above transcribed a particular song, clear sound quality (e.g., no excessive reverb) was prioritized.

After all 16 samples were generated, some quantitative measures were considered, such as the tempo and adjusted note density. When calculating the adjusted note density, "empty time" refers to periods of time with no melodic content, such as silence at the beginning of the prompt or periods in generated music containing non-piano instruments. The exact duration of empty time was determined by analyzing the audio waveform with free video editing software (e.g., Clipchamp).

$$\text{note density} = \frac{\text{note count}}{\text{total time}}$$

$$\text{adjusted note density} = \frac{\text{adjusted note density}}{\text{total time} - \text{empty time}}$$

We can also measure subjective accuracy based on others' perception of the generated music. After obtaining permission from the youth orchestra director, I distributed an online survey to students in a local youth orchestra via email. The form contained 16 samples in audio format as well as questions asking the recipient to determine the creator of the excerpt (human or AI), rate the excerpts' musicality on a Likert scale, and justify their rating. To prevent bias, the excerpts' order was randomized.

Survey responses were collected over a month (February 16, 2023 to March 12, 2023). The mean and standard deviation for both genres' ratings would immediately determine if AI or human music was preferred. A 2-sample t-test for a difference in means would determine whether there was a statistically significant difference between the ratings of human and AI-generated music.

Lastly, inductive thematic coding was applied to the free-response questions to see what positive and negative adjectives and phrases were used to describe human and AI-generated music. To do this, we analyzed the respondents' justifications for their ratings. Each response was coded by two independent individuals as strictly positive or negative; neutral responses were not counted. Insightful qualitative comments were included with commentary.

## 3 Results

In general, the generated samples had the same tempo and key signature as the prompts, showing that Jukebox had a basic understanding of musicality. The poor audio quality of some samples did not detract from this property. To best determine whether AI was able to accurately replicate rhythm game-style music, the research question has been split into four smaller questions below.

*1. How accurately did the MIDI data of the AI-generated samples (30-90 seconds) follow trends in the MIDI data of the human-generated prompts (0-30 seconds)?*

There were weak linear correlations between the tempo of the prompts and the adjusted note densities of the generated samples for both genres, artcore (r = -0.36) and orchestral (r = 0.12).

Therefore, a higher tempo did not correlate with a higher adjusted note density. This may be because the samples' original file format (.wav) had to be converted into MIDI. As such, the resulting noise made the adjusted note density impossible to predict, as seen from the presence of outliers in Figure 4.

**Figure 4**
(A) *Correlation between tempo of prompts and adjusted note densities of the 8 usable generated artcore samples.* (B) *Correlation between tempo of prompts and adjusted note densities of the 6 usable generated orchestral samples. Two of the eight samples were unable to be analyzed due to extremely long periods of non-melodic content such as drums, vocals, and non-piano instruments.*

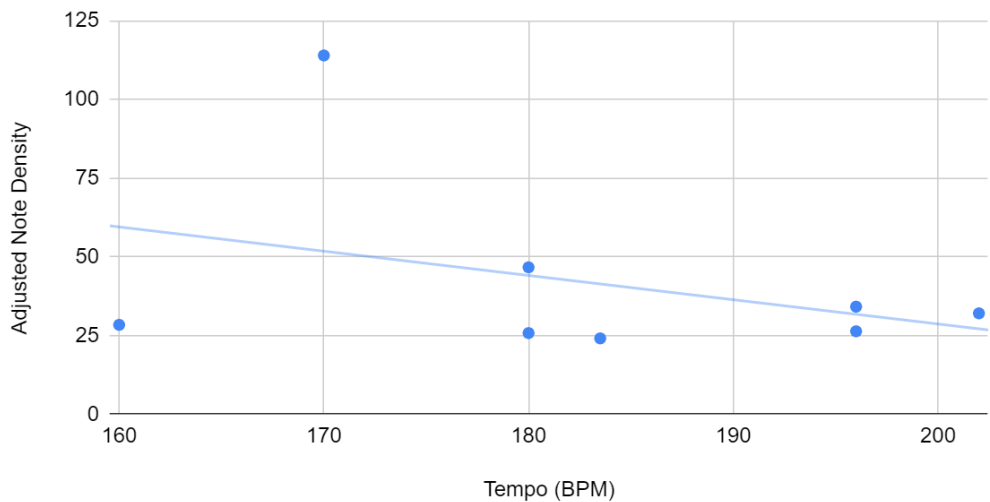

(A)

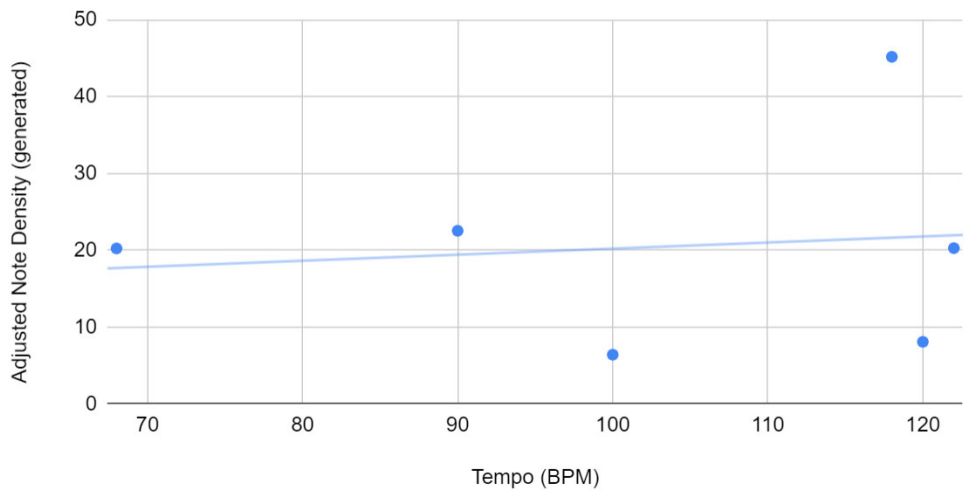

(B)

*2. Were the respondents able to determine the creator of each excerpt, human or AI?*

The survey utilized 8 human-generated excerpts and 8 AI-generated excerpts. According to the confusion matrix showed in Table 1, the respondents determined the creator correctly with 65.82% accuracy. For all AI-generated excerpts, the corresponding accuracy was 70.31%, and for all human-generated excerpts, the corresponding accuracy was 61.33%. The higher accuracy for AI-generated excerpts may be caused by artifacts in the audio files and usage of non-piano instruments, which caused respondents to be suspicious of their origin. Also, some human-generated compositions utilized unusual chord progressions or melodies, leading to misidentification. Other reasons are outlined in the discussion section.

**Table 1**
*Confusion matrix of answers to question 1.*

|  | Actually AI | Actually human | Total |
|---|---|---|---|
| Guessed AI | 180 | 99 | 279 |
| Guessed human | 76 | 157 | 233 |
| Total | 256 | 256 | 512 |

*Note:* The columns display the truth and the rows display the respondents' answers. Answers to different excerpts were considered distinct; there were 32 responses and thus 32 x 16 = 512 answers.

*3. Which excerpts were rated as more musical, human or AI?*

Music made by humans (mean = 3.83, SE = 0.97) was considered more musical than music made by AI (mean = 2.85, SE = 1.04). Figure 5 summarizes the results. Performing a 2-sample t-test for a difference of means with n = 32, the number of survey responses, yielded a t-value of 3.878 and a corresponding p-value of 0.000127. There is convincing statistical evidence that human-generated music was preferred.

**Figure 5**
*Comparison of mean ratings of human versus AI generated music from 1 to 5*

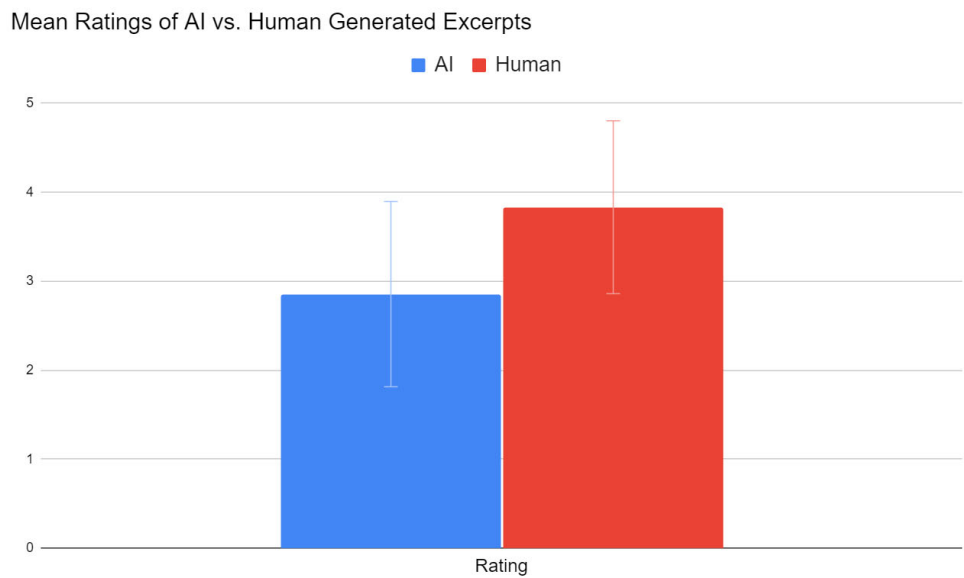

*4. If the respondents preferred certain excerpts over the others (e.g. human-generated excerpts), what were their reasons?*

Lastly, respondents tended to judge what they perceived to be AI-generated music more harshly than human-generated music when providing justifications for their ratings. Two aspects of each response were measured: one, whether the respondent judged the excerpt to be human or AI generated (regardless of correctness), and two, any positive or negative adjectives or phrases describing the excerpt. These qualitative responses were coded by two independent coders. The correlation coefficient between two coders was 0.955, which demonstrated that our decisions on the positive or negative feelings toward the music were consistent. These results have been summarized in Figure 6 below.

Human-generated music was often accompanied by positive adjectives. Respondents often used general adjectives (nice, cool, pretty, beautiful, etc.) in their descriptions. More importantly, respondents believed that these excerpts were structured and had direction; to some, the music sounded "familiar"; to others, it sounded "developed." There was a strong sentiment of musicality among the respondents, who noted logical chord progressions, phrasing, and layered melodies, among other aspects. One respondent cited specific examples of musicality from excerpt 5:

> It repeated familiar patterns and the chords sounded nice. although [sic] the glissando rolls were kind of repetitive, there was definite good structure with that rock solid 4-5-3-6 chord progression. The T-R [tension-resolution] seems very human and simple to understand, without mental gymnastics to justify them.

Such articulate justifications were rare, yet the mention of chord progressions and tension/resolution gave some insight as to what respondents listened for in an excerpt.

On the other hand, AI-generated music was accompanied by negative adjectives. Occurrences of general adjectives such as "bad" were rare, as respondents' criticisms were more specific. The corresponding negative adjectives were usually variants of random (chaotic), abrupt (sudden, jarring), and inhuman (unnatural, fake, "too perfect"). Lastly, just as human-generated music was praised as structured, AI-generated music was generally perceived as directionless and unstructured. Some of the responses may be caused by the excerpts' poor audio quality and/or presence of non-piano instruments; as one respondent notes about excerpt #9: "it sounds scary and the language is not anything that is spoken." [sic]

**Figure 6**
*Relative frequency of positive and negative adjectives in responses to question 3.*

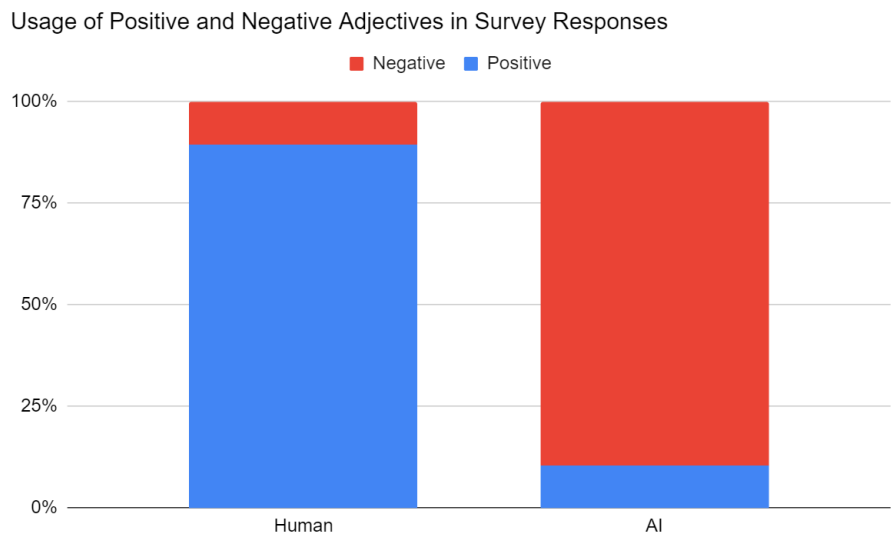

*Note*: On the left, the 161 positive and 19 negative adjectives correspond to excerpts respondents perceived as human. On the right, the 23 positive and 199 negative adjectives correspond to excerpts respondents perceived as AI-generated.

## 4  Discussion

To summarize the results, Jukebox could reasonably extend previously composed rhythm game music of two genres, artcore and orchestral, and mimic their styles. Even though survey respondents could generally tell whether a certain excerpt was composed by Jukebox, a mean rating of 2.85 signifies that some respondents viewed the AI-generated excerpts favorably.

Overall, the results of the current study confirm trends in AI-generated music found in other studies. For example, human-generated music tends to be rated more favorably (more musical) than AI-generated music. In a study by Pilat and Samuel (2019), respondents rated human-generated music 0.47 points higher than AI-generated music (compared to 0.98 in this study) with similar accuracy based on a dataset of over 100,000 MIDI files with no specific genre. Even though excerpts from both creators relied on MIDI, which would have eliminated glaring signs that an excerpt was AI-generated (such as poor audio quality and non-piano instruments, as stated previously), the difference can likely be attributed to AI not being able to extend musical patterns over large time scales. In another study performed by Chen et al. (2020), the researchers' neural network specifically designed to create multi-instrumental pop music was also successful in its goal, as it achieved a listener rating of 3.62 (compared to an AI rating of 2.85 in this study). However, it must be noted that the goal of Jukebox was not to produce multi-instrumental music. In all, these studies' results suggest that AI is capable of creating listenable music.

There were several major factors that impacted the results of this study that can be categorized as follows. Generating MIDI data proved difficult due to the nature of Jukebox, and the survey responses may have differed from expected.

*1. Generating MIDI data*

Firstly, the note density of the generated excerpts were much higher than expected because of Jukebox processing raw audio rather than MIDI. Nonresponse was a major issue when disseminating the survey, which resulted in less responses than expected. (However, the result that human-generated music was rated higher would be expected to stay the same.) Lastly, the excerpt lengths were limited, which made the response accuracy slightly higher than expected.

Various factors may have significantly impacted note density in the prompts and the generated excerpts. For example, prompts for trials 5 and 8 yielded unusually high note densities because the melodies included many 8th and 16th notes and the arrangers used chords judiciously. Furthermore, a song that used tremolo or glissandi (trials 6 and 8) increased the note density, as these musical features quantitatively require more notes to play. Because of this, a high tempo did not necessarily correlate with a high note density.

The generated excerpts were created with Jukebox via a Google Colab notebook due to a lack of experience with creating neural networks. One problem encountered early during the generation process was Colab's user limits. Colab lets a user borrow a GPU from one of Google's servers, and as such, Google has limited notebook runtime to 12 hours for free users. The laptop used to generate excerpts took approximately three hours to generate each sample, meaning all 16 samples were generated over the course of two weeks.

The MIDI data were converted from raw audio, meaning the MIDI files included many notes with low velocity and duration- "noise" left over from the conversion. Non-piano instruments such as vocals, drums, and strings also caused inflated note density in all excerpts. In extreme instances, drums and vocals persisted throughout the majority of the excerpt (#5 and #8) or the entire excerpt (#9 and #10), making the note density impossible to analyze for these excerpts.

*2. Gathering Survey Responses*

The next aspect of discussion is the survey. The questions were partly modeled after a previous study by Pilat and Samuel (2019) which involved multi-instrumental music. Questions 1 and 2 asked about distinguishing between human and AI-generated music and rating the quality of the music, respectively; as such, these questions were adapted into my survey. I added an optional third question asking respondents for their rating justifications, which would allow me to obtain the necessary information to answer the research question- the respondents' justifications for an excerpt's musicality or lack thereof.

Initially, there were a total of 32 questions; each of the 16 selected samples could be split into a prompt and a generated excerpt. This was cut down to 16 questions, including only 8 samples of each genre (artcore and orchestral). Despite this reduction, each question still took a minimum of 30 seconds to complete, meaning the entire survey would have to be completed in at least 10-15 minutes. The survey was also created on Google Forms, which does not natively support embedding audio. As a workaround, each question included a hyperlink to the audio file, yet in the case of bad internet connection, the audio file had to be downloaded. Lastly, the number of steps involved to complete the survey deterred many respondents from completing it. Thirty-two responses were gathered after emailing the youth orchestra members twice to encourage participation.

The relatively large standard deviation of the ratings can be partly attributed to the respondents' individual biases on what constituted human or AI generated music. For example, one respondent mistakenly believed that a human-generated sample was AI-generated because there was "no human error." Their assessment was correct, as all samples were generated based on MIDI arrangements of songs played by a computer, yet the actual composer of that particular sample was a human. Most surprisingly, a few respondents believed that some of the samples that heavily involved non-piano instruments (e.g., excerpt #9) were made by a human, despite the fact that all human-generated excerpts only included the piano.

Another factor that may have impacted on the overall accuracy of the responses was the excerpt lengths. Respondents only listened to the first 30 seconds of the generated excerpts. Jukebox's memory only takes into account the last few seconds of the audio it has listened to previously; therefore, it is difficult for it to detect and extrapolate musical patterns beyond this short time range. If respondents listened to the full 60 seconds of the generated excerpts as used to calculate note density, they would discover that Jukebox's output would no longer resemble the original prompt, whether that be suddenly changing keys in the middle of the excerpt or using chord progressions that did not resolve or sound "complete."

# 5 Conclusion

The goal of this study was to determine whether Jukebox, a neural network, was able to compose songs in two niche subgenres of music found in rhythm games: artcore and orchestral. The results have given convincing evidence that Jukebox has achieved this goal. Jukebox was limited to extending preexisting works and a narrow time frame of 60 seconds; however, the results show promise in neural networks' understanding of music and its structure in a genre of music not yet analyzed in the field.

## 5.1 Limitations

There were several limitations faced during research that were not listed in the discussion. To reiterate, Jukebox inputs and outputs raw audio, which makes its generated excerpts unpredictable; this was taken to the extreme when excerpts included non-piano instruments, such as percussion, strings, or voice. This was an aspect that could not be controlled and could impact the results by making some excerpts quantitatively impossible to analyze and making it easier for listeners to tell that an excerpt was generated by AI- possibly skewing the mean rating towards 1, less musical. This effect was mitigated through random selection of generated excerpts for the survey.

## 5.2 Future Directions

Future directions of this research could proceed in several different ways. To begin, researchers could use a neural network other than Jukebox that uses MIDI data as input. This would greatly reduce error associated with noise that results from using continuous data (values from 0-20 kHz) rather than discrete data (MIDI notes have integer values from 1-128). Discrete data would also allow for the calculation of the neural network's loss and for improved audio quality in the generated excerpts. Secondly, researchers could analyze more variables related to the generated music's ratings. Examples include excerpt length (longer excerpts might lead to lower ratings for AI-generated music due to its difficulty in extracting broad musical patterns), and respondent grouping (a certain age range or non-orchestra members would perhaps rate AI-generated music more highly than other groups of respondents). Lastly, access to greater computational resources would allow future researchers to generate excerpts of a greater length and quality than those in this survey. Respondents noted that poor audio quality made them suspicious that an excerpt was AI-generated, and improving the audio quality would likely make it more difficult to differentiate an excerpt's origin. Despite these limitations, this research nevertheless shows the promise of Jukebox as a capable composer. Future developments such as the ones listed above may further bridge the gap between AI and human-generated music.

## 6  Conflict of Interest

The authors declare that the research was conducted in the absence of any commercial or financial relationships that could be construed as a potential conflict of interest.

## 7  Author Contributions

NY: designed and authored the article; conceptualization and investigation; developed methods; collection and analysis of results; data curation and visualization.

## 8  Funding

There was no outside funding involved in the writing of this paper.

## 9  Acknowledgments

I would like to deeply thank Kelly Loveday for her guidance over the year this research was performed and Haotong Zhu for his insight. I would also like to thank Farragut High School for giving me the opportunity to conduct research in an academic setting.

## 11  Data Availability Statement

The raw data supporting the conclusions of this article will be made available by the authors, without undue reservation.